\documentclass[sigconf, nonacm]{acmart}
\usepackage{float}
\usepackage{tikz}
\usepackage{pgfplots}
\usepackage{stfloats}
\usepackage{hyperref}

\newcommand\vldbavailabilityurl{}
\newcommand\vldbpagestyle{plain}

\begin{document}
\title{IORM: Hierarchical I/O Governance for Thousands of Consolidated Databases on Oracle Exadata}

\author{Rajarshi Chowdhury}
\orcid{0009-0007-7032-2450}
\affiliation{%
  \institution{Oracle America Inc}
  \city{Redwood Shores}
  \state{CA}
  \postcode{94065}
}
\email{rajarshi.chowdhury@oracle.com}

\author{Akshay Shah}
\orcid{0009-0008-5049-9256}
\affiliation{%
  \institution{Oracle America Inc}
  \city{Redwood Shores}
  \state{CA}
  \postcode{94065}
}
\email{akshay.shah@oracle.com}

\author{Zakaria Alrmaih}
\affiliation{%
  \institution{Oracle America Inc}
  \city{Redwood Shores}
  \state{CA}
  \postcode{94065}
}
\email{zakaria.alrmaih@oracle.com}

\author{Chenhao Guo}
\affiliation{%
  \institution{Oracle America Inc}
  \city{Redwood Shores}
  \state{CA}
  \postcode{94065}
}
\email{chenhao.guo@oracle.com}

\author{Anubhav Singh}
\affiliation{%
  \institution{Oracle America Inc}
  \city{Redwood Shores}
  \state{CA}
  \postcode{94065}
}
\email{anubhav.a.singh@oracle.com}

\author{Sue Lee}
\affiliation{%
  \institution{Oracle America Inc}
  \city{Redwood Shores}
  \state{CA}
  \postcode{94065}
}
\email{sue.k.lee@oracle.com}
\begin{abstract}
Oracle Exadata consolidates thousands of tenant databases onto shared storage infrastructure deployed at hundreds of customer sites worldwide. Oracle Multitenant architecture enables this extreme density, with thousands of tenant databases sharing a single Exadata storage system---but this creates a multi-level resource hierarchy (container databases, tenant databases, and workloads within tenants) that commodity block-layer schedulers cannot govern, as they lack visibility into database semantics and tenant boundaries. This paper presents the I/O Resource Manager (IORM), a storage-side scheduler built on three mechanisms: \textit{I/O Tagging}, which propagates semantic context from the database kernel to the storage scheduler; \textit{Hierarchical Resource Profiles}, which express compositional allocation policies across consolidation tiers using shares and limits; and \textit{Unified Storage Governance}, which applies these policies consistently across all tiers of the storage hierarchy---persistent memory, flash, and hard disk---including cache placement decisions. IORM enables successful cloud deployments where thousands of tenants coexist on shared storage: production OLTP workloads run alongside concurrent analytical workloads from the same or different databases without noisy-neighbor interference. Evaluation on production Exadata systems demonstrates that IORM dramatically improves latency consistency, virtually eliminating tail latency outliers and delivering several-fold improvements in average read latency under mixed workloads. Hierarchical limits compose correctly across all three levels, and proportional share allocation tracks configured ratios closely even under highly skewed demand.
\end{abstract}

\maketitle

\pagestyle{\vldbpagestyle}
\begingroup
\renewcommand\thefootnote{}\footnote{\noindent
This is the authors' accepted manuscript. The final version will appear in the \emph{Proceedings of the VLDB Endowment} (PVLDB), 2026. \\
This work is licensed under the Creative Commons BY-NC-ND 4.0 International License. Visit \url{https://creativecommons.org/licenses/by-nc-nd/4.0/} to view a copy of this license. For any use beyond those covered by this license, obtain permission by emailing \href{mailto:info@vldb.org}{info@vldb.org}. Copyright is held by the owner/author(s). Publication rights licensed to the VLDB Endowment.
}\addtocounter{footnote}{-1}\endgroup

\ifdefempty{\vldbavailabilityurl}{}{
\vspace{.3cm}
\begingroup\small\noindent\raggedright\textbf{PVLDB Artifact Availability:}\\
The source code, data, and/or other artifacts have been made available at \url{\vldbavailabilityurl}.
\endgroup
}

\section{Introduction}
\label{sec:intro}

Database consolidation---running multiple databases on shared infrastructure---delivers compelling economic benefits but creates a fundamental resource governance challenge. When dozens or hundreds of tenant databases share common storage, how can the system ensure predictable performance for latency-sensitive workloads while allowing batch workloads to utilize available capacity?

Operating system I/O schedulers cannot solve this problem. Schedulers such as Linux CFQ, BFQ, and Kyber~\cite{cfq, bfq, kyber} operate at the block layer with no visibility into database semantics: to Linux, every 8KB read looks the same, whether it originates from a transaction commit blocking a user or a background maintenance task. Process-level controls like cgroups~\cite{cgroups} are equally ineffective because a single database process typically serves multiple tenant databases---the kernel cannot distinguish I/O from different tenants when it all comes from the same process. Hypervisor-level schedulers such as mClock~\cite{mclock} and PARDA~\cite{parda} operate at virtual machine boundaries, which similarly fail to capture the internal tenant structure.

This semantic gap has real consequences. Latency-sensitive operations such as transaction log writes lie on the critical path of commits; when these requests queue behind bulk scan I/Os, even brief delays cascade into application-visible latency spikes~\cite{tail-at-scale}. Prior work on storage QoS~\cite{argon, ioflow, cake} has demonstrated the importance of workload differentiation, yet existing solutions lack the semantic context to distinguish critical database operations from background maintenance at the granularity required for multi-tenant database consolidation.

This paper describes the I/O Resource Manager (IORM), a\break userspace storage-side scheduler deployed on Oracle Exadata~\cite{exadata} that addresses the semantic gap between database workloads and block I/O scheduling (Figure~\ref{fig:exadata-arch}). IORM runs in userspace on Exadata storage servers, where it intercepts I/O requests before they reach the physical devices. IORM's design rests on three key ideas.

First, I/O tagging: each I/O request carries metadata describing its semantic context---which tenant database issued it, what class of workload it belongs to, and what type of data it accesses. This tag travels with the request from the database to storage, enabling scheduling decisions that would be impossible at the block layer.

Second, hierarchical resource profiles: administrators express resource policies at multiple levels of a tenant hierarchy---container database, pluggable database, and workload. Policies compose hierarchically: a tenant cannot exceed its container's allocation regardless of how its internal workloads are configured. This enables independent administration at different organizational scopes while maintaining global constraints. Figure~\ref{fig:hierarchy} shows this resource hierarchy. The hierarchy is extensible; additional levels can be added as deployment complexity grows.

Third, unified storage governance: IORM applies its scheduling policies consistently across all tiers of the storage hierarchy---persistent memory, NVMe flash, and hard disk drives---as well as to cache placement decisions. Tags determine not only I/O dispatch priority but also which data should be promoted into flash cache, ensuring that cache resources are reserved for application data the database will re-access rather than being consumed by one-shot system traffic such as backup and storage rebalance.

Together, these mechanisms transform storage from a passive block device into an active participant in multi-tenant resource governance. The novelty of IORM lies not in any single primitive in isolation but in their combination at a particular architectural placement: storage-side semantic enforcement with hierarchical compositional policies across consolidation tiers. Section~\ref{sec:related} positions IORM in detail against prior storage QoS, hypervisor-side, and database-side governance work. This paper makes the following contributions:

\begin{itemize}
    \item We present the design and implementation of IORM, describing how I/O tagging propagates database semantic context from database servers to storage servers (Section~\ref{sec:tagging}), and how hierarchical resource profiles enable compositional policies across tenancy, and workload levels with support for shares and limits (Section~\ref{sec:profiles}). IORM governs all tiers of the storage hierarchy---persistent memory, flash, and hard disk---including cache placement, ensuring consistent policy enforcement across the entire storage stack.

    \item We detail the storage-side scheduling algorithms that enforce these policies, including lottery-based proportional sharing, cost-based utilization accounting, and deadline scheduling for starvation prevention (Section~\ref{sec:scheduling}).

    \item We evaluate IORM on production Exadata systems, demonstrating its effectiveness at protecting latency-sensitive workloads, enforcing proportional allocation, and preventing interference between tenants (Section~\ref{sec:evaluation}).

    \item We share operational lessons from deploying IORM across multiple Exadata hardware generations, including the design implications of hardware heterogeneity in long-lived storage pools (Section~\ref{sec:lessons}).
\end{itemize}  

\section{Background}
\label{sec:background}

\begin{figure}[h]
    \centering
    \includegraphics[width=0.95\columnwidth]{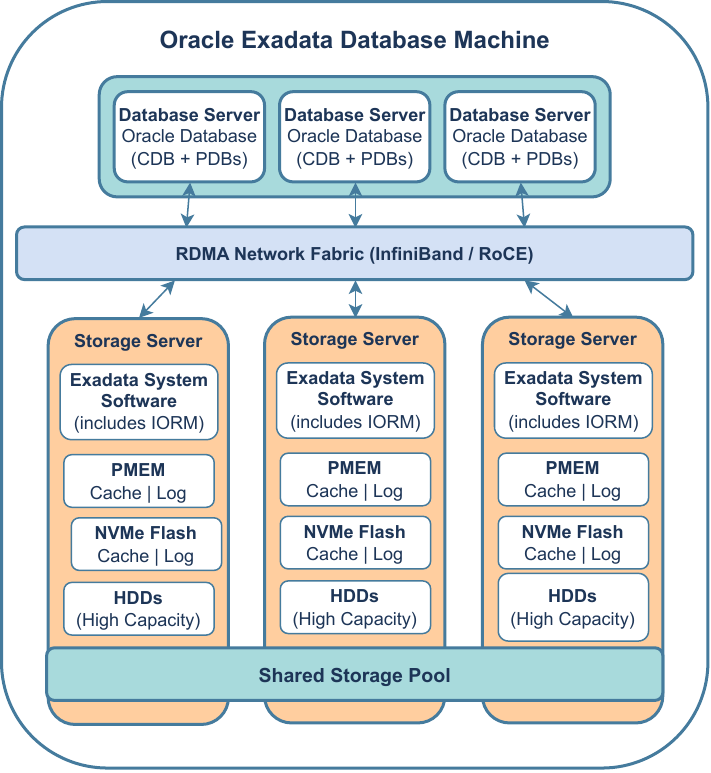}
    \caption{Oracle Exadata architecture. Database servers connect to intelligent storage servers over RDMA. Each storage server runs the IORM scheduler, which uses I/O tags to make semantic-aware scheduling decisions.}
    \Description{Diagram showing Oracle Exadata architecture with Database Servers connected to Storage Servers over RDMA. Each Storage Server runs the IORM scheduler.}
    \label{fig:exadata-arch}
\end{figure}

IORM works on Oracle Exadata Storage Servers. Oracle Exadata~\cite{exadata} is an engineered system combining database servers with intelligent storage servers connected via a high-speed RDMA network (InfiniBand or RoCE). Database servers run Oracle Database instances and issue I/O requests to storage servers, which manage the physical storage devices---hard disk drives, NVMe flash, and persistent memory. Figure~\ref{fig:exadata-arch} illustrates this architecture.

Unlike traditional storage arrays that present simple block devices, Exadata storage servers run software that can interpret\break database-level semantics. This capability is central to our approach: the storage layer can make scheduling decisions based on information that would be invisible to a generic block I/O scheduler.

\subsection{Multi-Tenant Database Consolidation}

Modern enterprises consolidate many databases onto shared infrastructure to improve utilization and reduce costs~\cite{multitenant}. Oracle's multi-tenant architecture supports this through a two-level hierarchy:

A \textit{Container Database} (CDB) is a database instance that can host multiple isolated tenant databases. The CDB provides shared infrastructure---memory, background processes, and common\break metadata---while maintaining strict isolation between tenants. In typical deployments, each CDB runs on its own database server or set of servers, while all CDBs share the same Exadata storage pool.

A \textit{Pluggable Database} (PDB) is a tenant database within a CDB. Each PDB appears to applications as an independent database with its own schemas, users, and data. PDBs can be created, cloned, and migrated independently, enabling agile provisioning. A single CDB can host up to 4096 PDBs.

Within each PDB, workloads are classified into \textit{PDB workloads}. A PDB workload groups database sessions with similar resource requirements---for example, interactive transactions, batch reports, or maintenance operations. Administrators assign sessions to workloads based on application identity, user credentials, or other criteria.

This creates a three-level resource hierarchy: CDB, PDB, and PDB workload. Figure~\ref{fig:hierarchy} illustrates this hierarchy.

\subsection{Production Deployment Scale}
\label{sec:deployment-scale}

On-premises deployments, Oracle Cloud Infrastructure, Exadata Cloud@Customer, multicloud deployments on AWS, Azure, and Google Cloud, Oracle Autonomous Database, and Exadata Database Service all run on the same Exadata substrate~\cite{exadata, exascale}. Oracle's 2025 Exadata Statement of Direction characterizes the platform as having tens of thousands of global deployments~\cite{exadata-sod}, and an independent industry analysis notes that Oracle disclosed more than 5{,}000 engineered-systems customers as of 2016 and has not released figures since, although the customer base is widely believed to span tens of thousands today~\cite{morgan-x11m}. Independent analyst coverage describes Exadata as a consolidation substrate for ``dozens, hundreds, thousands, and tens of thousands'' of Oracle databases of varied workload type on a single platform~\cite{cube-exadata}, and third-party customer-tracking databases catalog thousands of named Exadata and Exadata Cloud@Customer customers across industries~\cite{appsruntheworld-exadata, appsruntheworld-exacc}. Publicly disclosed scale points include Oracle Fusion Applications' Exadata Cloud Service fleet of approximately 20{,}000 databases and the Swiss telecommunications operator Sunrise's reported migration of roughly 1{,}000 databases to Exadata Cloud@Customer~\cite{sunrise-migration}. Oracle Exascale~\cite{exascale, cw-exascale}, the disaggregated-storage successor architecture, is also designed around the IORM concepts described in this paper. The IORM mechanisms described in this paper are the in-production governance layer that makes such consolidation density tractable: every Exadata storage server in every one of these deployments runs the IORM scheduler, and every I/O issued by an Oracle Database session against shared Exadata storage is governed by it.

\subsection{The Scheduling Gap}

The multi-tenant hierarchy described above is invisible to conventional I/O schedulers. Block-layer schedulers see only offsets and sizes; they cannot determine which tenant issued a request or whether it is latency-critical. Process-level isolation fails because all tenants within a CDB share database processes. Hypervisor-level schedulers~\cite{mclock, parda} can enforce VM-level policies but cannot see the tenant or workload structure within each VM.
Without visibility into this hierarchy, administrators face an unpleasant choice: either accept unpredictable interference between tenants, or abandon consolidation by dedicating storage to individual databases~\cite{ioflow, cake}. IORM resolves this dilemma by propagating database semantics to the storage layer, enabling fine-grained governance without sacrificing consolidation density.

\section{I/O Tagging}
\label{sec:tagging}

The key insight enabling database-aware storage scheduling is that the database engine possesses semantic information unavailable to external schedulers. IORM exploits this through \textit{I/O tagging}: each I/O request carries metadata describing its origin and purpose, enabling the storage layer to make informed scheduling decisions.

\subsection{Tag Contents}
\label{sec:tag-contents}

Every I/O request from the database to storage carries a compact tag containing:

\begin{itemize}
    \item Tenant identity: which PDB (and CDB) issued the request. This enables the storage scheduler to enforce per-tenant resource policies.
    
    \item PDB workload: which workload the request belongs to (e.g., interactive, batch, maintenance). This enables differentiated service within a tenant.
    
    \item I/O category: whether the request is for user data, transaction log, temporary data, or internal metadata. Certain categories---particularly transaction log writes---are latency-critical because they lie on the commit path~\cite{tail-at-scale}. Table~\ref{tab:io-types} summarizes the I/O types and their characteristics.
    
    \item Priority hint: an explicit priority designation.\break System-initiated I/O (such as backup or reorganization tasks) typically carries low priority to avoid impacting user-facing workloads.
\end{itemize}

Categories and priorities are independent dimensions because the same I/O type can have very different urgency depending on \textit{why} it is being issued. Category encodes \textit{what} the I/O accesses (a static property of the data), while priority encodes \textit{how urgent} this particular request is (a dynamic property of the issuing operation). Redo log writes, for example, normally carry high priority because they are on the commit path. Redo log reads issued for media recovery during instance restart and redo archival reads issued by a background log-shipper share the same category but warrant very different scheduling treatment. Decoupling the two dimensions lets administrators express policy in natural terms (``cap maintenance traffic to 5\% regardless of which file types it touches'') without expanding the category enumeration.

The database kernel populates these fields from information already available in memory: the session's tenant context, its assigned workload, and the file type being accessed. The priority field is set by the issuing subsystem according to its role (log writer, database writer, recovery, backup, parallel query coordinator). Tag generation adds negligible overhead---under 100 nanoseconds per I/O. Prior work on semantic-aware storage~\cite{semantically-smart} has demonstrated the value of application context, but to our knowledge, IORM is the first production storage scheduler to implement database-level tagging at this consolidation scale. The same tagging approach is portable to any disaggregated database architecture such as Aurora, Neon, or AlloyDB by extending the storage RPC protocol to carry equivalent metadata.

Beyond scheduling, I/O tags also govern flash cache placement. Exadata's flash cache is a shared resource, and without semantic awareness it would be vulnerable to pollution from one-shot system traffic that has no expectation of being re-read, most notably backup and storage rebalance, which scan large volumes of data on behalf of system-level operations rather than application queries. IORM uses the I/O category tag (Table~\ref{tab:io-types}) to determine cache eligibility: I/O that the database is likely to re-access (including both latency-sensitive buffer cache reads and analytic scan blocks over user data) is admitted to the cache, while background operations such as backup and storage rebalance bypass the cache entirely. This prevents cache pollution from one-shot maintenance traffic and ensures that cache capacity is reserved for the data the application actually re-reads.

\begin{table}[t]
    \centering
    \footnotesize
    \setlength{\tabcolsep}{3pt}
    \caption{I/O types generated by Oracle Database with varying latency requirements. Tags encode this semantic information to enable both prioritization and cache placement decisions.}
    \begin{tabular}{llll}
        \toprule
        \textbf{I/O Type} & \textbf{Pattern} & \textbf{Latency Sensitivity} & \textbf{Cache} \\
        \midrule
        Redo log writes & Seq., small & Critical (commit) & Write-back \\
        Buffer cache reads & Rand., small & High (queries) & Yes \\
        Direct path reads & Seq., large & Medium (scans) & Conditional \\
        Database writes & Rand., small, batch & Low (checkpoint) & Write-back \\
        Temp writes & Seq., large & Medium (sorts) & No \\
        Undo writes & Rand., small & Medium (rollback) & Yes \\
        Backup & Seq., large & Low (background) & No \\
        Storage rebalance & Mixed & Low (maintenance) & No \\
        \bottomrule
    \end{tabular}
    \label{tab:io-types}
\end{table}

\subsection{Tag Propagation}

Tags travel with I/O requests across the network from database servers to storage servers. Exadata uses a protocol layered on RDMA that includes the tag in each request header. When the request arrives at the storage server, the scheduler extracts the tag and uses it to determine queue placement and dispatch priority.

Concretely, the tag is carried as a small, fixed-format buffer adjacent to each I/O request rather than embedded in the data block payload. It consists of a versioned header followed by a payload identifying the issuing tenant (database identifier) and the physical target (file number, block offset, block count), smaller than 64\,KB in total. Multi-byte fields are stored in network byte order so that tags generated on database servers can be interpreted by storage servers of any architecture. The higher-level classifications described in Section~\ref{sec:tag-contents} (workload class, I/O category, and priority) are not transmitted as separate fields. The storage server derives them at request arrival by looking up the tag's primitives in metadata provisioned at PDB creation time. Keeping the wire format compact and policy-agnostic lets categories and policies evolve on the storage side without protocol changes.

Integration into the database kernel is modest in scope. The issuing subsystem (log writer, database writer, recovery, backup, parallel query coordinator) constructs the tag at request creation time, since it already holds both the physical addressing and the operation type. The request-submission path attaches the tag as an explicit parameter on the outgoing RPC rather than relying on thread-local storage or implicit context propagation. The same approach is portable to other disaggregated database architectures such as Aurora, Neon, and AlloyDB: the requirements are an RPC-style storage protocol that can carry per-I/O metadata, a storage-side scheduler that interprets the tag, and a tag-construction step in each I/O-issuing subsystem of the database.

For I/O that passes through intermediate layers (such as a volume manager), the tag is preserved and forwarded. This ensures that the storage scheduler sees the original semantic context regardless of how many software layers the request traverses.

\subsection{Handling Untagged I/O}

Not all I/O carries tags. Requests from non-database applications, operating system utilities, or older database versions may lack tagging. The storage scheduler assigns such requests to a default category with baseline resource allocation. This ensures that untagged workloads receive service but cannot benefit from semantic prioritization.

\begin{figure*}[b]
    \centering
    \includegraphics[width=1.9\columnwidth]{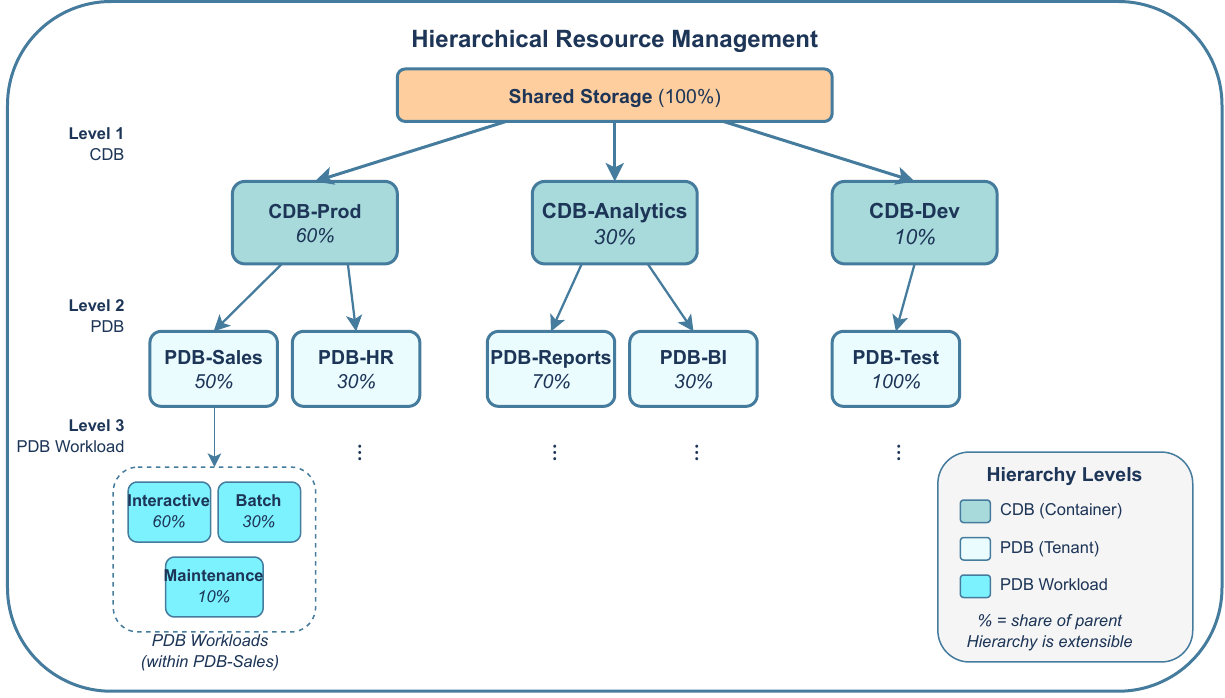}
    \caption{Resource management hierarchy in consolidated Exadata environments. Policies at each level constrain the resources available to lower levels: CDB $\rightarrow$ PDB $\rightarrow$ PDB Workload. The hierarchy is extensible to additional levels.}
    \Description{Diagram showing the resource management hierarchy in consolidated Exadata environments, with policies constraining resources at each level.}
    \label{fig:hierarchy}
\end{figure*}

\section{Hierarchical Resource Profiles}
\label{sec:profiles}

I/O tags provide the semantic context for scheduling decisions; resource profiles define the policies that govern how resources are allocated. IORM uses a hierarchical profile framework where policies at each level constrain the resources available to lower levels.

\subsection{The Resource Hierarchy}

IORM organizes resource governance into three levels:

\begin{enumerate}
    \item CDB level: each CDB receives a share of total storage capacity. This level is relevant when multiple CDBs share common storage.
    
    \item PDB level: each PDB within a CDB receives a share of its container's allocation.
    
    \item PDB workload level: each workload within a PDB receives a share of its tenant's allocation.
\end{enumerate}

The hierarchy is extensible. Deployments requiring finer-grained control can add levels above (e.g., for multi-cluster environments) or below (e.g., for sub-workload classification) as needed.

The key property of this hierarchy is \textit{compositional isolation}: a tenant cannot exceed its container's allocation regardless of how its workloads are configured. An administrator managing a PDB can freely adjust workload policies without affecting other tenants or violating CDB-level constraints. Figure~\ref{fig:hierarchy} illustrates this hierarchy with example allocations.

\subsection{Allocation Mechanisms}

At each level, administrators can specify resource allocations using two mechanisms:

Shares provide proportional allocation. If tenant A has 3 shares and tenant B has 1 share, A receives 75\% of available bandwidth when both are active. Shares are \textit{work-conserving}: when B is idle, A receives 100\%. Unused capacity automatically flows to active workloads. This approach follows the proportional-share scheduling model used in hypervisor schedulers~\cite{mclock, waldspurger-vmware}.

Limits impose hard caps. A tenant with a 40\% limit cannot exceed that utilization even if the system is otherwise idle. Limits are useful for preventing runaway workloads from monopolizing resources, or for enforcing ``pay for performance'' contracts where tenants should not exceed their purchased capacity.

These mechanisms can be combined. A common pattern is ``share among available capacity, limit to 60\%''---allowing bursting when capacity is available, but preventing monopolization. This flexibility exceeds what is available in most storage QoS systems~\cite{ioflow, cake}.

The examples above express allocations as percentages of storage capacity. Percentages are the natural unit for proportional sharing because they are \textit{hardware-independent}: a 25\% share means the same thing whether the underlying devices are hard disks or NVMe flash. Internally, IORM translates these percentage-based policies into concrete I/O rate limits using a hardware-aware cost model (Section~\ref{sec:lessons} discusses the operational motivation for this design).

The same percentage-based policy is applied uniformly across HDD, NVMe flash, and PMEM rather than exposed as per-tier knobs. The database, not the administrator, decides which tier serves any given I/O: PMEM and flash are managed as caches and write-staging tiers transparently to the application, so a tenant cannot meaningfully target ``20\% of NVMe and 5\% of HDD'' because it does not control which tier its requests land on. The right unit of policy is therefore the tenant's overall I/O entitlement, not its per-device entitlement, and per-tier limits would penalize cache-resident tenants by forcing artificial throttling at higher tiers. Exposing per-tier control would also force administrators to reason jointly about working-set behavior, cache eligibility, and tier headroom, which is intractable at consolidation densities of hundreds to thousands of PDBs per system. For the rare cases where an administrator does need to influence tier behavior, IORM exposes coarse-grained controls, most importantly the cache-eligibility decisions driven by I/O category tags (Section~\ref{sec:tagging}), rather than per-tier rate limits.

\subsection{Policy Composition}

When multiple hierarchy levels are active, IORM computes an \textit{effective allocation} for each I/O request by composing policies across levels. Consider a request from workload \texttt{BATCH} in tenant \texttt{PDB-Sales} within container \texttt{CDB-Prod}:

\begin{enumerate}
    \item The CDB-level profile specifies \texttt{CDB-Prod} receives 60\% of storage capacity.
    \item The PDB-level profile specifies \texttt{PDB-Sales} receives 25\% of \texttt{CDB-Prod}'s allocation---effectively 15\% of total capacity.
    \item The workload-level profile specifies \texttt{BATCH} receives 20\% of \texttt{PDB-Sales}'s allocation---effectively 3\% of total capacity.
\end{enumerate}

This multiplicative composition ensures that lower-level policies cannot circumvent higher-level constraints. The storage scheduler uses these effective allocations when making dispatch decisions.

\subsection{Administrative Model}

The hierarchy maps naturally to organizational roles: storage administrators configure CDB-level policies, database administrators configure PDB-level policies, and application administrators configure workload-level policies using Oracle Database Resource Manager~\cite{dbrm}. Each administrator operates independently within their scope---a DBA can add new PDBs or adjust tenant shares without involving the storage administrator, as long as the aggregate stays within the container's allocation. This separation of concerns scales to large organizations with distributed administrative responsibility.

\subsection{Scheduling Objective}

Beyond per-entity allocations, IORM provides a global \textit{objective} that influences scheduling behavior. Four modes are available: \textit{low latency} (shallow queues, prioritize small I/Os), \textit{high throughput} (deeper queues, favor large sequential I/Os), \textit{balanced} (trade off between latency and throughput), and \textit{auto} (detect workload characteristics and adapt automatically). Auto is the default and recommended setting. The objective provides a system-wide tuning knob that complements per-entity profiles, similar to the approach taken by feedback-based resource controllers~\cite{feedback-control, padala-eurosys}. The objective applies uniformly across the deployment and is intended as a coarse, infrequently-changed setting (most production deployments leave it at \textit{auto}). Per-tenant differentiation is expressed through profiles, not through the objective.

\section{Storage-Side Scheduling}
\label{sec:scheduling}

The IORM scheduler on each storage server receives tagged I/O requests and dispatches them to storage devices according to profile policies and workload characteristics. The scheduler must balance latency protection for critical I/Os~\cite{tail-at-scale}, proportional sharing under contention~\cite{mclock}, strict limit enforcement, work conservation, and starvation prevention~\cite{argon}. This section describes how IORM achieves these goals.

\subsection{Queue Depth Control}

The fundamental design principle behind IORM's device-level scheduling is to issue only a limited number of I/Os to each storage device at any time, so that the scheduler retains control over the order in which requests are serviced. If the scheduler allowed hundreds of I/Os to accumulate in a device queue, it would lose the ability to reorder or preempt requests---the device would process them in whatever order its internal firmware chose. By keeping the device queue shallow, IORM ensures that when a high-priority request arrives, it can be dispatched promptly rather than waiting behind a backlog of lower-priority traffic.

For hard disks, IORM targets a steady-state read queue depth of 62 outstanding I/Os per spindle, partitioned so that small (latency-sensitive) requests have a \emph{guaranteed floor} of 32 slots while large I/Os are capped at no more than 10 concurrent in-flight requests. Small I/Os can grow into the remaining read-queue capacity above their 32-slot floor when large traffic is light; large I/Os never exceed their cap. Writes are governed by a separate threshold of 8 outstanding requests, independent of the 62-slot read budget. This partitioning balances utilization (allowing throughput-oriented scans to keep the disk busy) against latency (preventing large I/Os or write bursts from filling the queue and blocking small requests). Without queue depth control, a burst of low-priority scan traffic could fill the device queue with hundreds of requests, causing subsequent high-priority requests to wait seconds for service.

Large I/Os consume more queue capacity than small I/Os---a 1MB read occupies the disk longer than an 8KB read. The scheduler accounts for this by weighting large I/Os at 3x the cost of small I/Os when computing queue depth.

Queue depth is enforced at I/O dispatch time as a token-style admission check. Each device maintains a counter of in-flight cost using the same weighting, and the scheduler will dispatch a request from its queues only if doing so would not push that counter above the configured target (62 for HDD reads in steady state with the small/large partition described above, 8 for HDD writes, higher and dynamic for flash, see Section~\ref{sec:device-specific}). When an I/O completes, the device emits a completion event, the in-flight cost is decremented by the request's cost, and the scheduler is signalled to re-evaluate dispatch. The target itself can move at runtime: when flash cache is unavailable (for example during cache warm-up or while the cache tier is being serviced), the HDD target is reduced from 62 to 32 so that workloads forced to fall through to disk do not queue up against the deeper steady-state target. This makes the device queue a dynamic, self-clocking structure: enforcement happens implicitly because no new I/O is dispatched until completions free up cost, rather than via explicit polling or rate-limiting timers.

\subsection{Lottery Scheduling}

When queue capacity becomes available, the scheduler selects the next request using lottery scheduling weighted by share allocations~\cite{lottery-scheduling}. Each entity with queued I/O holds ``tickets'' proportional to its share. The scheduler draws a ticket to select which entity may issue next.

The lottery operates hierarchically: first selecting a CDB, then a PDB within that CDB, then a workload within that PDB. At each level, selection probability follows share proportions. This ensures that hierarchical share allocations compose correctly.

Entities that have reached their utilization limits are excluded from the lottery until the next accounting interval. This prevents limited workloads from consuming more than their allocation even when they have queued requests.

\subsection{Utilization Accounting}

To enforce limits, the scheduler tracks each entity's utilization over time. The challenge is defining ``utilization'' in a meaningful way: raw I/O counts are misleading because a 1MB read consumes more device capacity than an 8KB read.

IORM uses \textit{cost-based accounting}. The cost of an I/O is the time the device spends servicing it, excluding queue wait time. The scheduler computes this by measuring device busy time and I/O completions over one-second intervals. An entity's utilization is its consumed cost divided by total available cost.

Accounting operates at two granularities: 200-millisecond \textit{quanta} for responsive limit enforcement, and 1-second intervals for accurate accounting~\cite{rate-limiting}. At each quantum boundary, the scheduler compares actual utilization against limits and temporarily throttles entities that have exceeded their allocation. At each 1-second boundary, the scheduler reconciles across quanta, carrying forward small deficits or surpluses.

The reconciliation step exists because quantum-level decisions are inherently noisy: a 200ms window contains a small number of large I/Os, so a single in-flight 1MB request that completes just after the window boundary can swing measured utilization by several percentage points. If each quantum were enforced independently, a lightly-loaded tenant whose I/Os happened to land at quantum boundaries would see oscillating throttling. The 1-second reconciliation treats the five quanta in each second as a single accounting frame: at the end of the second, the scheduler computes the per-entity actual-versus-budget delta and updates a per-entity carry-forward credit, clamped to a small bound (a few percentage points). The carry-forward credit then biases the next second's quantum-level enforcement: an under-consumed entity gets a slightly higher effective limit in the next second, allowing it to recover briefly delayed I/O. An over-consumed entity operates under a slightly tighter effective limit until the credit is repaid. The clamp prevents an idle entity from accumulating arbitrary credit and bursting in a way that would itself violate the limit. This two-tier scheme delivers responsive throttling at the 200ms time scale while preserving accurate proportional enforcement at the 1-second time scale.

\subsection{Deadline Scheduling}

Pure priority scheduling risks starvation: if high-priority traffic is continuous, low-priority requests might wait indefinitely. IORM prevents this through deadline scheduling.

Each queued request is timestamped on arrival. A background thread periodically scans for requests that have waited longer than a threshold---1 second by default. Requests exceeding this deadline are promoted to a ``starved'' list that the scheduler checks before running the normal lottery.

This mechanism guarantees that every request not currently subject to active limit throttling completes within bounded time, regardless of its original priority, while still providing strong differentiation under normal conditions.

One might ask why the deadline mechanism is uniform rather than priority-graded: once a request crosses the 1-second deadline, it is promoted ahead of all regular lottery selection, regardless of original priority. We considered priority-graded alternatives, such as promoting low-priority starved requests to a ``mid'' tier rather than the front of the line, or scaling the deadline threshold inversely with priority, but rejected them. The deadline is failure-mode insurance: if a request has waited a full second on a storage device, something is operationally abnormal (severe over-subscription, runaway high-priority workload, hardware degradation), and the right response is to bound worst-case latency rather than preserve fine-grained priority ordering at the tail. A 1-second deadline is roughly 1{,}000$\times$ a normal flash service time, so by the time it triggers, normal scheduling has already failed for that request. In a healthy system the mechanism almost never fires, so a graded promotion scheme would buy very little under normal load while making corner-case behavior harder to reason about. We prefer the simpler invariant: ``no request waits more than one second.'' An important corollary is that the deadline mechanism does not override hard limits: even when a request crosses the deadline, it is not promoted past limit enforcement. A tenant that has exhausted its utilization cap remains excluded until the next accounting interval restores eligibility (Section~\ref{sec:profiles}). Limits and deadlines are deliberately ranked: limits express contractual entitlements that must hold regardless of load, while the deadline is a best-effort starvation guard that yields to those entitlements. Frequent firing in production is a signal the deployment is undersized or misconfigured, not that the deadline policy needs tuning. The scheduler exposes deadline-promotion counts as real-time statistics, and storage administrators routinely configure alerts on sustained non-zero firing rates as an early indicator that capacity, workload behavior, or hardware health requires investigation.

\subsection{Workload-Adaptive Behavior}

The scheduler detects workload characteristics and adapts its behavior:

When only one workload is active (solo workload), the scheduler disables throttling and allows maximum device concurrency. There is no contention to manage, so governance overhead provides no benefit.

When a high-priority workload issues predominantly small I/Os (latency-sensitive priority workload), indicating interactive traffic, the scheduler restricts low-priority traffic to minimal concurrency and fragments large low-priority I/Os into smaller chunks. Fragmentation prevents a single 1MB scan read from blocking interactive requests for the full duration of a large sequential transfer.

When active workloads are predominantly large-I/O (throughput-oriented workload), indicating batch or scan traffic, the scheduler increases device concurrency to maximize bandwidth.

A dedicated thread evaluates workload patterns every few seconds and sets the scheduling mode accordingly.

\subsection{Device-Specific Adaptations}
\label{sec:device-specific}

The mechanisms described above---queue depth control, lottery scheduling, utilization accounting, deadline scheduling, and workload adaptation---apply across device types, but flash and hard disk storage require different parameterizations.

\textbf{Flash.} Service times are orders of magnitude lower than hard disks~\cite{ssd-tradeoffs, nvme-latency}, so queuing overhead is proportionally significant. IORM therefore bypasses queuing entirely for high-priority I/Os (system-critical operations, small user reads, latency-sensitive workloads), dispatching them immediately. Only low-priority I/Os---large scans and background operations---pass through the scheduler's queues and throttling logic, with a default target of 8 outstanding low-priority I/Os per device.

\textbf{Hard disk.} Mechanical seek times and rotational latency make queue management critical. IORM maintains a per-spindle read queue depth target of 62 outstanding I/Os in steady state, with 32 of those slots guaranteed as a floor for small (latency-sensitive) requests and large I/Os capped at no more than 10 concurrent in-flight requests; small I/Os can occupy unused read-queue capacity above their 32-slot floor, preventing unbounded device queue growth. When a latency-sensitive OLTP workload coexists with scans, IORM fragments large scan I/Os into 128KB chunks, allowing small interactive I/Os to interleave rather than stalling behind multi-millisecond transfers. Exadata storage controllers include supercapacitor-backed write cache, so cached writes complete in under 1ms; IORM uses a separate threshold of 8 outstanding writes, independent of the read budget, and monitors cache pressure to pace low-priority I/O when the controller must flush to disk.

Exadata storage spans both flash and hard disk: flash itself is offered in both TLC NVMe (faster, higher endurance) and QLC NVMe (denser, lower cost-per-TB) configurations, and high-capacity deployments combine these with high-density hard disks. The HDD tier remains a first-class part of the system because of capacity economics: customer datasets are routinely in the hundreds of terabytes per database and the multiple-petabyte range per consolidated deployment, and at those sizes the per-TB cost gap between HDD and even QLC flash is wide enough to make HDD the rational choice for the bulk of the data footprint. Exadata's design exploits the typical access skew in OLTP and analytical workloads: a relatively small hot working set drives most of the I/O, fits in flash cache, and is served at NVMe latencies, while older partitions, archival rows, and infrequently-touched data live on HDD and are touched only when queries reach into the cold tier. This tiered design is what allows Exadata to deliver flash-level latency for the active working set without paying the per-terabyte cost of an all-flash configuration, a value proposition that has only grown more pronounced as enterprise NAND prices have risen sharply under AI-driven demand for SSD capacity~\cite{trendforce-nand-2026}. The IORM hard-disk code paths govern this cold tier in essentially every high-capacity deployment and remain the dominant I/O path for full-table scans of large historical tables, backup and restore traffic, and Smart Scan offload over data that has aged out of cache. The noisy-neighbor results in Section~\ref{sec:eval-noisy} (Figure~\ref{fig:latency-histogram}) are deliberately measured on the HDD path because that is where queueing is most pronounced and where uncontrolled bulk I/O does the most damage to interactive latency.

\section{Evaluation}
\label{sec:evaluation}

We evaluate IORM on production Exadata X8-2 systems using controlled experiments that isolate specific scheduling properties. In production, IORM governs deployments consolidating hundreds to thousands of PDBs per Exadata system, where typical storage utilization ranges from 30--70\% and the scheduler processes on the order of millions of I/O decisions per second per storage cell. The controlled testbed below uses smaller configurations to enable precise measurement; the scheduling algorithms and code paths are identical to production.

We measure effectiveness across three dimensions: protecting latency-sensitive workloads from noisy neighbors, enforcing hierarchical resource policies, and achieving proportional share allocation under contention.

\subsection{Noisy Neighbor Protection}
\label{sec:eval-noisy}

The ``noisy neighbor'' problem---where a bandwidth-intensive workload degrades performance for latency-sensitive tenants---is the primary challenge IORM addresses. We evaluate this on an Exadata X8-2 Half Rack with 4 compute nodes and 6 storage cells, configuring two PDBs within a shared CDB.

Both PDBs use the standard IORM intra-database directives so that policy enforcement reflects what a typical production deployment would configure. The \textit{OLTP PDB} runs the OLTP workload with all sessions in an \textit{interactive} workload class tagged as latency-sensitive (high priority). The \textit{Scan PDB} runs the analytical workload with all sessions in a \textit{batch} workload class tagged as throughput-oriented (low priority). At the PDB level, both PDBs are given equal shares. No hard limits are configured, so all governance flows from share-weighted scheduling and priority-aware queue management. This configuration deliberately stresses the scheduler rather than relying on hard caps to do the work. The system-wide objective is set to \textit{auto}. Profiles are identical in the IORM-enabled and IORM-disabled runs. The only variable is whether the IORM scheduler is active or whether I/O is dispatched via the bypass path used as the disabled baseline.

The OLTP workload simulates interactive transaction processing: 200 concurrent sessions performing single-block random reads (8KB) against tables containing 256 million rows each. Each session executes point queries representing user-facing operations such as order lookup or account inquiry. The dataset fits entirely in flash cache, so under ideal conditions reads complete in sub-millisecond time.

The analytical workload simulates decision support: a hash join across two large tables from a 10TB dataset, using degree of parallelism 300 across 3 compute nodes. This query generates large sequential I/Os and spills approximately 1TB to temporary storage---representative of batch reporting or business intelligence dashboards.

\begin{table}[h]
    \centering
    \small
    \caption{OLTP throughput (reads/sec) when running alone versus concurrently with an analytical workload. IORM reduces throughput degradation by 2.6$\times$, recovering a substantial portion of the performance lost to noisy-neighbor interference.}
    \begin{tabular}{lrrr}
        \toprule
        \textbf{Configuration} & \textbf{OLTP Alone} & \textbf{OLTP Mixed} & \textbf{Degradation} \\
        \midrule
        IORM Disabled & 468k/s & 200k/s & $-$57\% \\
        IORM Enabled & 441k/s & 345k/s & $-$22\% \\
        \bottomrule
    \end{tabular}
    \label{tab:oltp-protection}
\end{table}

Table~\ref{tab:oltp-protection} shows the impact. Without IORM, large sequential I/Os from the analytical query flood device queues, causing interactive requests to stall behind bulk traffic and cutting OLTP throughput by more than half. With IORM enabled, I/O tags allow the scheduler to identify and prioritize latency-sensitive traffic, recovering a substantial portion of the lost throughput---a 2.6$\times$ reduction in degradation.
\begin{figure}[h]
    \centering
    \begin{tikzpicture}
        \begin{axis}[
            ybar,
            width=\columnwidth,
            height=5.5cm,
            bar width=6pt,
            ylabel={Percentage of Reads (\%)},
            xlabel={Latency},
            ymin=0,
            ymax=80,
            xtick=data,
            xticklabels={$<$512$\mu$s, $<$1ms, $<$2ms, $<$4ms, $<$8ms, $<$16ms, $<$32ms, $\geq$32ms},
            xticklabel style={font=\small, rotate=45, anchor=east},
            ylabel style={font=\small},
            xlabel style={font=\small},
            legend style={
                at={(0.97,0.97)},
                anchor=north east,
                font=\small,
                draw=none,
                fill=white,
                fill opacity=0.8,
                text opacity=1,
            },
            legend cell align={left},
            enlarge x limits=0.08,
            grid=major,
            grid style={dashed, gray!30},
            ymajorgrids=true,
            xmajorgrids=false,
        ]
        
        \addplot[
            fill=gray!60,
            draw=gray!80,
        ] coordinates {
            (1, 62.2)
            (2, 23.2)
            (3, 8.9)
            (4, 3.2)
            (5, 1.2)
            (6, 0.5)
            (7, 0.4)
            (8, 0.5)
        };
        
        \addplot[
            fill=blue!60,
            draw=blue!80,
        ] coordinates {
            (1, 50.0)
            (2, 44.3)
            (3, 5.4)
            (4, 0.2)
            (5, 0.1)
            (6, 0.0)
            (7, 0.0)
            (8, 0.0)
        };
        
        \legend{IORM Disabled, IORM Enabled}
        
        \end{axis}
    \end{tikzpicture}
    \caption{Single-block read latency distribution on hard disk storage under mixed workload. Without IORM, 0.5\% of reads exceed 32ms with outliers reaching 1 second. With IORM, 99.7\% of reads complete within 4ms---tail latency outliers are reduced below operationally significant thresholds.}
    \label{fig:latency-histogram}
\end{figure}
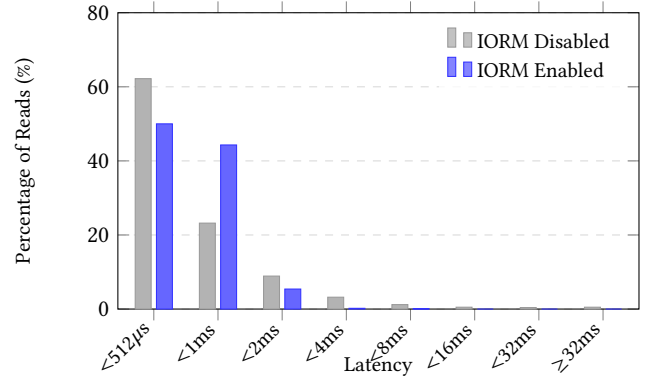
The most striking improvement, however, is in latency consistency. Figure~\ref{fig:latency-histogram} shows the single-block read latency distribution on hard disk storage under the mixed workload. Without IORM, the distribution has a long tail with outliers reaching 1 second---stalls immediately visible to end users. With IORM, the tail is eliminated entirely: 99.7\% of reads complete within 4ms and average latency improves 4.5$\times$ (from 960$\mu$s to 213$\mu$s). This tail latency elimination is the most operationally significant result: in production deployments, it is the unpredictable latency spikes---not average throughput reduction---that generate customer complaints and violate application SLAs.

\begin{table}[h]
    \centering
    \small
    \caption{Per-PDB IORM queue time on flash under mixed OLTP+Analytics load, as the per-device queue-depth target is swept. The OLTP PDB stays at nanosecond-scale queue time across all configurations. The Scan PDB absorbs the queueing.}
    \begin{tabular}{lrrr}
        \toprule
        \textbf{Queue depth target} & \textbf{OLTP PDB} & \textbf{Scan PDB} & \textbf{Ratio} \\
        \midrule
        8 (default) & 20\,ns & 3.72\,ms & $\sim$186{,}000$\times$ \\
        16 & 23\,ns & 53.9\,$\mu$s & $\sim$2{,}300$\times$ \\
        32 & 23\,ns & 3.96\,$\mu$s & $\sim$170$\times$ \\
        64 & 12\,ns & 607\,ns & $\sim$50$\times$ \\
        \bottomrule
    \end{tabular}
    \label{tab:per-pdb-queue}
\end{table}

To complement the HDD latency-distribution view, we also measured per-PDB IORM queue time on the flash tier of the same Exadata system under an analogous mixed-workload pattern, with the cache-resident OLTP working set isolated to flash so that all OLTP reads are served from flash cache. The flash tier does not benefit from the latency-histogram view (cache-resident reads complete in microseconds regardless of contention, so there is no tail to eliminate), but it does expose a different facet of tag-based scheduling: how queueing pressure is partitioned between PDBs. Table~\ref{tab:per-pdb-queue} reports the IORM queue time observed at the OLTP PDB and the Scan PDB across a sweep of the per-device queue-depth target. The OLTP PDB sees IORM queue times in the tens of nanoseconds across every configuration. Tag-driven priority routes its requests around scan-induced queueing entirely, while the Scan PDB absorbs the queueing pressure, with its queue time falling by orders of magnitude as the queue-depth target rises. The four-to-five orders-of-magnitude gap between the two PDBs is the direct signature of tag-based scheduling: the same physical device serves both, but the latency-sensitive tenant is held flat at near-zero queue time regardless of how the throughput-oriented tenant is configured.

\begin{table*}[b]
    \centering
    \small
    \caption{Hierarchical limit enforcement across CDB and PDB levels. Utilization values are flash device utilization (\%). Higher-level limits strictly constrain lower levels. Effective PDB utilization respects $\min(\text{PDB limit}, \text{CDB limit})$.}
    \label{tab:limit-enforcement}
    \begin{tabular}{clccccc}
        \toprule
        \textbf{Case} & \textbf{Policy Configuration} & \textbf{CDB-A} & \textbf{CDB-B} & \textbf{Key PDB Utils (A)} & \textbf{Key PDB Utils (B)} & \textbf{Observation} \\
        \midrule
        1 & No limits (baseline) & 33.2\% & 33.3\% & PDB5: 23.1\% & PDB8: 22.6\% & Equal sharing \\
        2 & CDB-A limit=10\% & 10.0\% & 60.4\% & PDB5: 7.2\% & PDB8: 45.1\% & CDB limit enforced \\
        3 & CDB-A limit=10\%, PDB5 limit=20\% & 10.0\% & 63.2\% & PDB5: 2.0\% & PDB8: 46.9\% & CDB cap binds first \\
        4 & PDB5 limit=1\% (no CDB limit) & 14.5\% & 52.5\% & PDB5: 1.0\% & PDB8: 40.5\% & PDB limit enforced \\
        5 & CDB-A limit=1\%, PDB5 limit=10\% & 1.0\% & 80.2\% & PDB5: 0.1\% & PDB8: 61.9\% & Tight CDB overrides PDB \\
        6 & CDB-A limit=1\%, PDB5 limit=1\% & 1.0\% & 71.1\% & PDB5: $<$0.01\% & PDB8: 53.5\% & Both levels enforced \\
        \bottomrule
    \end{tabular}
\end{table*}

In summary, IORM's queue depth management and semantic-aware scheduling deliver dramatic improvements under mixed workloads: average latency improves 4.5$\times$, throughput degradation is reduced by 2.6$\times$, and tail latency outliers are eliminated entirely. The analytical workload also benefits---large I/O queue time drops from 54ms to 15ms (a 3.6$\times$ improvement)---because IORM prevents any single workload from saturating the device queue. On the flash tier, where the latency distribution is dominated by cache hits and offers little headroom for tail-latency improvement, the per-PDB queue-time breakdown (Table~\ref{tab:per-pdb-queue}) confirms that tag-based scheduling differentiates between latency-sensitive and throughput-oriented tenants by orders of magnitude on the same shared device.

\subsection{Hierarchical Limit Enforcement}
\label{sec:eval-limits}

A core property of IORM's hierarchical profiles is \textit{compositional isolation}: limits at higher levels constrain all entities below, regardless of lower-level policies. We validate this across the full three-level hierarchy---CDB, PDB, and PDB workload---with 10 configurations exercising different limit combinations.

We configure two CDBs sharing an Exadata storage pool, each containing 5~PDBs. PDBs within CDB-A are PDB1--5; PDBs within CDB-B are PDB6--10. We drive concurrent small I/O workloads across selected PDBs and measure per-entity flash device utilization.

Table~\ref{tab:limit-enforcement} summarizes six representative configurations. Three key observations confirm correct hierarchical enforcement. First, CDB-level limits are enforced precisely: capping CDB-A at 10\% (Case~2) frees capacity that flows to CDB-B, confirming work conservation. Second, hierarchical composition is correct: in Case~3, PDB5's 20\% limit is overridden by its container's tighter 10\% cap, while in Case~4 a PDB-level limit constrains only the targeted tenant without affecting siblings. Third, tight cascading limits compose as expected: when CDB-A is limited to 1\% (Case~5), all its PDBs share that narrow budget regardless of their individual settings.

Table~\ref{tab:three-level-limits} extends the evaluation to three active hierarchy levels: CDB, PDB, and PDB workload. Each PDB's sessions run in a workload with a 10\% limit.

\begin{table}[h]
    \centering
    \small
    \caption{Three-level limit enforcement: CDB $\rightarrow$ PDB $\rightarrow$ PDB workload. Workload limit = 10\% for all PDBs. Effective utilization respects the tightest constraint at any level.}
    \label{tab:three-level-limits}
    \begin{tabular}{lccc}
        \toprule
        \textbf{Policy} & \textbf{CDB-A} & \textbf{PDB1 (wkld)} & \textbf{PDB5 (wkld)} \\
        \midrule
        CDB-A lim=50\%, PDB5 lim=5\% & 10.2\% & 5.0\% & 0.2\% \\
        CDB-A lim=10\%, PDB5 lim=1\% & 2.0\% & 1.0\% & 0.01\% \\
        CDB-A lim=1\%, PDB5 lim=1\% & 0.2\% & 0.1\% & 0.01\% \\
        No CDB/PDB limits & 22.5\% & 6.3\% & 10.0\% \\
        \bottomrule
    \end{tabular}
\end{table}

The results confirm that limits compose multiplicatively across all three levels. The binding constraint shifts depending on which level is tightest: when the CDB cap is generous, workload-level limits dominate; when the CDB cap tightens, it becomes the binding constraint and workload utilization drops proportionally. The minimum enforceable utilization is 0.01\% at the PDB level.

\subsection{Proportional Share Allocation}
\label{sec:eval-shares}

Shares govern proportional allocation when multiple entities contend for capacity. Unlike limits, shares are work-conserving: they determine how \textit{contested} capacity is divided but do not restrict utilization when capacity is available.

We configure two CDBs (CDB-A, CDB-B) each running 200 concurrent large I/O operations, generating sufficient load to create queueing at the storage devices. We vary the share ratio between CDBs across four configurations.

\begin{table}[h]
    \centering
    \small
    \caption{Proportional share allocation between two CDBs under contention. Share ratios translate proportionally to throughput and utilization, and inversely to I/O wait time.}
    \label{tab:share-allocation}
    \begin{tabular}{lrrrr}
        \toprule
        \textbf{Share Ratio} & \textbf{CDB-A} & \textbf{CDB-B} & \textbf{Actual} & \textbf{CDB-A} \\
        \textbf{(A : B)} & \textbf{(MB/s)} & \textbf{(MB/s)} & \textbf{Ratio} & \textbf{Wait (ms)} \\
        \midrule
        1 : 1 (baseline) & 10,330 & 10,323 & 1.00 & 17.1 \\
        2 : 1 & 13,817 & 6,919 & 2.00 & 12.1 \\
        5 : 1 & 17,177 & 3,520 & 4.88 & 9.1 \\
        10 : 1 & 18,785 & 1,989 & 9.44 & 8.0 \\
        \bottomrule
    \end{tabular}
\end{table}

Table~\ref{tab:share-allocation} shows the results. Throughput tracks share proportions closely across all ratios, with wait times mirroring the distribution inversely. The slight deviation from ideal at 10:1 (actual ratio 9.44$\times$) reflects scheduling quantization effects in the 200ms accounting window. Wait times for the disfavored CDB rise to 99.6ms at the most skewed configuration, confirming that shares redistribute both bandwidth and queueing delay.

We also verify that share changes take effect promptly under load. While I/O is running, we switch from a (CDB-A=2, CDB-B=1) share plan to (CDB-A=1, CDB-B=2). Within one minute, the throughput ratio inverts: CDB-A drops from 13,742 to 6,895~MB/s while CDB-B rises from 6,884 to 13,749~MB/s. The symmetric results (within 0.2\%) confirm that the scheduler rebalances cleanly without hysteresis.

\subsection{Scheduling Overhead}
\label{sec:eval-overhead}

IORM's scheduling decisions add latency to the I/O path. We quantify this overhead on an idle system by comparing I/O latency with IORM enabled versus bypassed. For 8KB random reads on flash, IORM adds less than 5$\mu$s of scheduling overhead per I/O---negligible relative to typical device service times of 80--200$\mu$s. Tag extraction and lottery selection together consume under 2$\mu$s; the remainder is queue management bookkeeping. CPU utilization attributable to IORM on each storage cell is below 3\% of a single core even under peak I/O load, leaving ample headroom for storage server processing. The workload-adaptive behavior (Section~\ref{sec:scheduling}) further reduces overhead: when only a single workload is active, the scheduler bypasses throttling and lottery selection entirely.

\subsection{Cache Governance}
\label{sec:eval-cache}

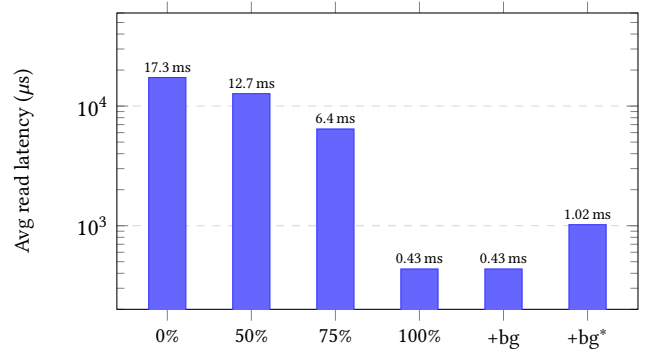
\begin{figure}[h]
    \centering
    \begin{tikzpicture}
        \begin{axis}[
            ybar,
            ymode=log,
            log basis y=10,
            width=\columnwidth,
            height=5.5cm,
            bar width=14pt,
            ylabel={Avg read latency ($\mu$s)},
            ymin=200, ymax=60000,
            symbolic x coords={0\%, 50\%, 75\%, 100\%, +bg, +bg$^{*}$},
            xtick=data,
            xticklabel style={font=\small},
            ylabel style={font=\small},
            point meta=explicit symbolic,
            nodes near coords,
            nodes near coords style={font=\scriptsize, anchor=south, inner sep=2pt},
            enlarge x limits=0.12,
            ymajorgrids=true,
            grid style={dashed, gray!30},
        ]
        \addplot[fill=blue!60, draw=blue!80] coordinates {
            (0\%,   17315) [17.3\,ms]
            (50\%,  12696) [12.7\,ms]
            (75\%,  6438)  [6.4\,ms]
            (100\%, 435)   [0.43\,ms]
            (+bg,    435)   [0.43\,ms]
            (+bg$^{*}$, 1020) [1.02\,ms]
        };
        \end{axis}
    \end{tikzpicture}
    \caption{Average read latency for the latency-sensitive workload (log scale). Bars 1--4 sweep the PDB's cache quota with no other workload running. \textit{+bg} adds a concurrent background workload under the default tag policy (bypasses cache). \textit{+bg$^{*}$} repeats with the cache-exclusion tag disabled.}
    \label{fig:cache-governance}
\end{figure}

We quantify the value of tag-aware cache placement with two complementary measurements on the same Exadata X8-2 testbed used in Section~\ref{sec:eval-noisy}. First, we sweep the flash cache quota allocated to a single latency-sensitive PDB issuing 8KB random reads against a fixed working set, from no cache to full coverage, with no other workload running. Second, we hold the latency-sensitive PDB at full cache coverage and run a concurrent background workload tagged as cache-ineligible (per Table~\ref{tab:io-types}), comparing the default tag policy against a modified policy in which the same background traffic is admitted to the cache. Figure~\ref{fig:cache-governance} reports average read latency for the latency-sensitive workload across all six configurations. The cache-size sweep confirms that latency is dominated by cache coverage: average latency falls from 17.3\,ms with no cache to 0.43\,ms when the working set fits in flash. The background-workload comparison directly measures the incremental benefit of tag-aware cache placement, with scheduling held constant. Under the default policy, the latency-sensitive workload retains its fully-cached latency (0.43\,ms) because the background traffic bypasses cache entirely. With the cache-exclusion tag disabled, background blocks displace cached data, average latency rises 2.4$\times$ to 1.02\,ms, and the latency distribution widens by an order of magnitude (read-latency standard deviation grows from 228\,$\mu$s to 3.87\,ms). The cost of the default policy is that the background workload's throughput drops by roughly 4$\times$ when forced to disk, an operationally acceptable trade for traffic whose latency does not appear on the application path.

\section{Operational Lessons}
\label{sec:lessons}

As Exadata scaled to support cloud customers, we encountered a fundamental challenge: long-lived storage pools are rarely homogeneous. Rolling hardware upgrades mean that a single deployment may contain storage servers spanning multiple generations---for example, X8 and X10 nodes in the same pool---with raw I/O throughput varying by 2--3$\times$. A tenant limited to 100,000 IOPS would consume a much larger fraction of capacity on an older server than on a newer one, leading to uneven throttling and unpredictable performance. This drove a fundamental design decision: IORM expresses all resource accounting in \textit{utilization percentages} rather than raw I/O rates. A 10\% limit means 10\% of whatever the underlying hardware can deliver, regardless of generation. The scheduler translates between percentages and absolute rates using a hardware-aware cost model normalized to each device's rated capacity. When a storage server is replaced with a newer generation, all tenant policies continue to work without reconfiguration. For cloud deployments where tenants are provisioned with explicit IOPS entitlements, the scheduler performs the inverse translation---converting the IOPS target into a utilization percentage relative to the specific hardware serving the request.

In cloud deployments, tenants are provisioned with an IOPS entitlement---for example, 32,000 IOPS for a block volume---and they expect this number to represent a \textit{consistent floor}, not a best-case peak. If a tenant observes 32,000 IOPS during quiet periods but only 20,000 under contention, the provisioned limit has failed its purpose. At the same time, cloud economics depend on overprovisioning: the aggregate provisioned IOPS across all tenants typically exceeds physical capacity, since not all tenants are active simultaneously. This creates a tension---the system must guarantee each tenant's floor while gracefully handling the inevitable moments when too many tenants are active at once. IORM resolves this through the combination of limits (which cap individual tenants to their entitlement during contention), shares (which distribute remaining capacity proportionally), and deadline scheduling (which prevents starvation). Without this active governance, overprovisioned storage pools would suffer unpredictable performance under load spikes, undermining the SLA commitments that cloud customers depend on.

Perhaps the most surprising lesson is that most customers evaluate storage performance not by their provisioned IOPS but by the amount of flash cache space available to their databases. A customer whose working set fits in flash cache experiences sub-millisecond latency and rarely encounters IOPS limits; a customer whose working set exceeds the cache sees frequent disk I/Os with much higher latency, and perceives ``slow storage'' even when their IOPS entitlement is fully delivered. IORM's tag-based cache placement (Table~\ref{tab:io-types}) is therefore as important in practice as its I/O scheduling: ensuring that application data the database will re-read remains cached, and that one-shot maintenance traffic such as backup and storage rebalance does not displace it---often matters more to end-user experience than the precise IOPS allocation. This motivated the unified governance approach where tags control both scheduling priority and cache eligibility.

\subsection{Policy Tuning in Practice}
\label{sec:policy-tuning}

A common question from production DBAs is how shares and limits should be chosen, how often they should be revisited, and what kinds of misconfiguration we have seen.

Most deployments start by translating the commercial agreement into IORM policy: a tenant entitled to 32{,}000 IOPS gets a corresponding limit, and shares within a CDB default to equal weights unless the consolidation plan explicitly assigns relative priorities (e.g., production gets 4 shares, development gets 1). DBAs almost never start from a measured workload profile because measured profiles are not yet available at provisioning time. They start from the contract and tune against observed behavior afterward. The work-conserving nature of shares means under-tuned configurations rarely cause problems while spare hardware capacity exists.

Adjustment cadence is event-driven rather than periodic. Policies are revisited not on a schedule but in response to specific events: onboarding of a new tenant, a recurring noisy-neighbor incident in monitoring, a tier-mix change in the underlying hardware, or renegotiation of a tenant's commercial entitlement. Steady-state production deployments often run for months without policy changes. The hierarchical model supports this: a DBA can adjust a workload-level share for one PDB without coordinating with the storage administrator, as long as the change stays within the PDB's container budget.

Three patterns account for most of the misconfiguration support tickets we see. The first is conflating share with limit: administrators set a 25\% share intending to cap a tenant, then are surprised when the tenant uses 90\% as the only active workload. Shares are work-conserving by design. A hard cap requires a limit. The second is cascading limits that compose to near-zero: a CDB cap of 10\% combined with a PDB cap of 10\% combined with a workload cap of 10\% yields an effective 0.1\% budget. We address this by surfacing the effective multiplicative budget in monitoring tools rather than just per-level percentages. The third is relying on shares to enforce SLOs: shares determine how contested capacity is divided, so a tenant accustomed to a high share will see latency degrade only when contention appears. SLO enforcement requires a combination of shares and limits, and the ``share fairly, but cap at 60\%'' idiom of Section~\ref{sec:profiles} exists precisely to make this expressible.

\section{Related Work}
\label{sec:related}

Operating system I/O schedulers have evolved from simple elevator algorithms to proportional-share designs. Linux CFQ~\cite{cfq} introduced per-process queues with time-slice fairness; BFQ~\cite{bfq} improved responsiveness with budget-based scheduling; Kyber~\cite{kyber} targets NVMe SSDs with a lightweight two-queue design. These schedulers operate at the block layer and differentiate traffic by process or cgroup, which is insufficient when a single database process serves multiple tenants. IORM provides a semantic-aware layer above block scheduling.

Substantial prior work addresses storage QoS. Argon~\cite{argon} pioneered performance insulation using interposed scheduling.\break Fa\c{c}ade~\cite{facade} introduced virtual storage devices with performance guarantees, and Triage~\cite{triage} applied adaptive control for differentiated service. PARDA~\cite{parda} and mClock~\cite{mclock} addressed proportional allocation in virtualized environments---mClock's tag-based algorithm supporting reservations, limits, and shares influenced IORM's allocation primitives. IOFlow~\cite{ioflow} proposed software-defined storage with centralized policy and distributed enforcement, while Cake~\cite{cake} demonstrated SLO enforcement through two-level scheduling. Pisces~\cite{pisces-storage} explored isolation for multi-tenant cloud storage, and ReFlex~\cite{reflex} provided predictable remote flash access. These systems influenced IORM's design but do not address the deep multi-tenant hierarchy (CDB/PDB/workload) specific to consolidated database environments. Three properties together distinguish IORM from this body of work. First, \textit{depth of hierarchy}: prior systems generally expose a flat tenant model or at most two levels (e.g., Cake's two-level scheduling, mClock's per-VM allocation), whereas IORM composes policies multiplicatively across three or more levels (CDB, PDB, workload) corresponding to distinct administrative scopes. Second, \textit{database-level semantics}: prior storage QoS systems differentiate traffic by client, VM, or flow identifier, which is too coarse when a single database process serves many tenants and many I/O classes. IORM propagates tenant identity, workload class, and I/O category from the database kernel down to the storage scheduler. Third, \textit{unified governance across scheduling and cache placement}: prior systems treat I/O dispatch and cache admission as separate concerns, whereas IORM uses the same tag to drive both, ensuring that bulk traffic does not monopolize device queues and that one-shot maintenance traffic does not displace application data from flash cache.

The idea that storage can benefit from application context dates to work on semantically-smart disk systems~\cite{semantically-smart}, which showed that file-system hints improve scheduling and caching. IORM extends this to database workloads, propagating tenant identity, workload classification, and I/O type to enable database-specific scheduling at the scale of thousands of tenants.

Database resource managers such as Oracle Database Resource Manager~\cite{dbrm} govern CPU and I/O within database instances. IORM extends this model to storage, providing consistent resource management across the stack. At the cluster level, Borg~\cite{borg} and Heracles~\cite{heracles} demonstrated that latency-critical and batch workloads can share infrastructure with careful management. IORM addresses similar challenges in storage, protecting OLTP latency from batch interference. The proportional-share scheduling underlying IORM's allocation was formalized in lottery scheduling~\cite{lottery-scheduling} and applied to hypervisor memory management~\cite{waldspurger-vmware}; IORM adapts these techniques to hierarchical I/O scheduling with variable I/O costs and heterogeneous devices.

Cloud-native databases such as Amazon Aurora, Google Spanner, and CockroachDB internalize I/O scheduling within the query engine or rely on cloud-provider storage abstractions (e.g., EBS, Colossus) for isolation. These architectures sidestep the multi-tenant storage scheduling problem by coupling each database tightly to its storage layer. IORM addresses a complementary scenario: consolidating many existing Oracle databases---often migrated from dedicated infrastructure---onto shared Exadata storage without requiring application changes. The storage-side approach preserves compatibility with the full Oracle Database ecosystem while providing governance guarantees comparable to dedicated deployments.

\section{Conclusion}
\label{sec:conclusion}
The central lesson from building and deploying IORM is that storage
scheduling for consolidated databases is fundamentally a semantic
problem, not a throughput problem. Block-layer schedulers fail not
because they lack sophistication, but because they lack context: without knowing which tenant issued a request or whether it lies on a commit path, no scheduling algorithm can make the right trade-off. I/O tagging closes this gap by making the storage layer aware of information that was previously confined to the database kernel.

The hierarchical profile model proved equally important in practice.
Consolidation involves multiple administrative domains---storage teams, DBAs, and application owners---each needing independent control within global constraints. Compositional isolation makes this tractable: administrators at each level operate freely without risk of violating policies set above them.

Looking ahead, the extensible hierarchy provides a foundation for
governance beyond the current three tiers. As deployment topologies
grow---spanning multiple clusters, regions, or cloud tenancy
models---additional levels can be inserted without redesigning the
scheduling framework. Oracle Exascale~\cite{exascale}, which expands Exadata's capabilities as cloud storage platform, is built on the same IORM concepts described in this paper but deepens the hierarchy further: it extends governance to disaggregated storage pools shared across compute and block-volume workloads, adds distributed feedback controllers to resolve access skew across storage nodes, and introduces hardware-aware cost models that normalize IOPS accounting across heterogeneous device generations.

\begin{acks}
We thank the members of the Oracle Exadata, Exascale, and Resource Management teams for their contributions to the design, implementation, and deployment of IORM.
\end{acks}

\balance
\bibliographystyle{ACM-Reference-Format}
\bibliography{sample}

@inproceedings{multitenant,
  author    = {{Oracle Corporation}},
  title     = {Best Practices for Database Consolidation in {Oracle} Multitenant},
  booktitle = {Oracle White Paper},
  year      = {2023},
  note      = {\url{https://www.oracle.com/a/tech/docs/database/maa-consolidation.pdf}}
}

@misc{exadata,
  author    = {{Oracle Corporation}},
  title     = {Oracle Exadata Database Machine Technical Overview},
  year      = {2024},
  howpublished = {Oracle Technical White Paper},
  note      = {\url{https://www.oracle.com/engineered-systems/exadata/}}
}

@misc{cfq,
  author       = {Jens Axboe},
  title        = {{CFQ} {I/O} Scheduler},
  howpublished = {Linux Kernel Documentation},
  note         = {\url{https://www.kernel.org/doc/Documentation/block/cfq-iosched.txt}}
}

@inproceedings{bfq,
  author = {Valente, Paolo and Andreolini, Mauro},
  title = {Improving Application Responsiveness with the {BFQ} Disk {I/O} Scheduler},
  year = {2012},
  publisher = {Association for Computing Machinery},
  doi = {10.1145/2367589.2367590},
  booktitle = {Proceedings of the 5th Annual International Systems and Storage Conference (SYSTOR)},
  articleno = {6},
  numpages = {12},
  location = {Haifa, Israel}
}

@inproceedings{kyber,
  author = {Ren, Zebin and Doekemeijer, Krijn and Trivedi, Animesh},
  title = {A Systematic Configuration Space Exploration of the {Linux} {Kyber} {I/O} Scheduler},
  year = {2024},
  publisher = {Association for Computing Machinery},
  doi = {10.1145/3629527.3651416},
  booktitle = {Companion of the 15th ACM/SPEC International Conference on Performance Engineering (ICPE)},
  pages = {167--173},
  location = {London, United Kingdom}
}

@article{cgroups,
  author = {Down, Chris},
  title = {5 Years of Cgroup v2: The Future of {Linux} Resource Control},
  journal = {;login: The USENIX Magazine},
  volume = {46},
  number = {2},
  year = {2021},
  publisher = {USENIX Association}
}

@inproceedings{mclock,
  author = {Gulati, Ajay and Merchant, Arif and Varman, Peter J.},
  title = {{mClock}: Handling Throughput Variability for Hypervisor {I/O} Scheduling},
  booktitle = {9th USENIX Symposium on Operating Systems Design and Implementation (OSDI)},
  year = {2010},
  pages = {437--450},
  address = {Vancouver, BC},
  publisher = {USENIX Association}
}

@inproceedings{parda,
  author = {Gulati, Ajay and Ahmad, Irfan and Waldspurger, Carl A.},
  title = {{PARDA}: Proportional Allocation of Resources for Distributed Storage Access},
  booktitle = {7th USENIX Conference on File and Storage Technologies (FAST)},
  year = {2009},
  pages = {85--98},
  address = {San Francisco, CA},
  publisher = {USENIX Association}
}

@article{tail-at-scale,
  author = {Dean, Jeffrey and Barroso, Luiz Andr\'{e}},
  title = {The Tail at Scale},
  year = {2013},
  volume = {56},
  number = {2},
  doi = {10.1145/2408776.2408794},
  journal = {Communications of the ACM},
  pages = {74--80}
}

@inproceedings{argon,
  author = {Wachs, Matthew and Abd-El-Malek, Michael and Thereska, Eno and Ganger, Gregory R.},
  title = {Argon: Performance Insulation for Shared Storage Servers},
  booktitle = {5th USENIX Conference on File and Storage Technologies (FAST)},
  year = {2007},
  pages = {61--76},
  address = {San Jose, CA},
  publisher = {USENIX Association}
}

@inproceedings{ioflow,
  author = {Thereska, Eno and Ballani, Hitesh and O'Shea, Greg and Karagiannis, Thomas and Rowstron, Antony and Talpey, Tom and Black, Richard and Zhu, Timothy},
  title = {{IOFlow}: A Software-Defined Storage Architecture},
  year = {2013},
  publisher = {Association for Computing Machinery},
  doi = {10.1145/2517349.2522723},
  booktitle = {Proceedings of the 24th ACM Symposium on Operating Systems Principles (SOSP)},
  pages = {182--196},
  location = {Farmington, PA}
}

@inproceedings{cake,
  author = {Wang, Andrew and Venkataraman, Shivaram and Alspaugh, Sara and Katz, Randy and Stoica, Ion},
  title = {Cake: Enabling High-Level {SLOs} on Shared Storage Systems},
  year = {2012},
  publisher = {Association for Computing Machinery},
  doi = {10.1145/2391229.2391243},
  booktitle = {Proceedings of the 3rd ACM Symposium on Cloud Computing (SoCC)},
  articleno = {14},
  location = {San Jose, CA}
}

@misc{dbrm,
  author    = {{Oracle Corporation}},
  title     = {Managing Resources with {Oracle Database Resource Manager}},
  year      = {2024},
  howpublished = {Oracle Database Administrator's Guide},
  note      = {\url{https://docs.oracle.com/en/database/oracle/oracle-database/23/admin/managing-resources-with-oracle-database-resource-manager.html}}
}

@inproceedings{ssd-tradeoffs,
  author    = {Agrawal, Nitin and Prabhakaran, Vijayan and Wobber, Ted and Davis, John D. and Manasse, Mark and Panigrahy, Rina},
  title     = {Design Tradeoffs for {SSD} Performance},
  booktitle = {USENIX Annual Technical Conference (ATC)},
  year      = {2008},
  pages     = {57--70},
  publisher = {USENIX Association}
}

@inproceedings{nvme-latency,
  author    = {Li, Huaicheng and Hao, Mingzhe and Tong, Michael Hao and Sundararaman, Swaminathan and Bjorling, Matias and Gunawi, Haryadi S.},
  title     = {The {CASE} of {FEMU}: Cheap, Accurate, Scalable and Extensible Flash Emulator},
  booktitle = {17th USENIX Conference on File and Storage Technologies (FAST)},
  year      = {2018},
  pages     = {83--90},
  publisher = {USENIX Association}
}

@book{feedback-control,
  author    = {Hellerstein, Joseph L. and Diao, Yixin and Parekh, Sujay and Tilbury, Dawn M.},
  title     = {Feedback Control of Computing Systems},
  publisher = {John Wiley \& Sons},
  year      = {2004}
}

@inproceedings{padala-eurosys,
  author    = {Padala, Pradeep and Shin, Kang G. and Zhu, Xiaoyun and Uysal, Mustafa and Wang, Zhikui and Singhal, Sharad and Merchant, Arif and Salem, Kenneth},
  title     = {Automated Control of Multiple Virtualized Resources},
  booktitle = {Proceedings of the 4th ACM European Conference on Computer Systems (EuroSys)},
  pages     = {13--26},
  year      = {2009},
  address   = {Nuremberg, Germany},
  publisher = {ACM}
}

@inproceedings{rate-limiting,
  author    = {Raghavan, Barath and Vishwanath, Kashi and Ramabhadran, Sriram and Yocum, Kenneth and Snoeren, Alex C.},
  title     = {Cloud Control with Distributed Rate Limiting},
  booktitle = {Proceedings of ACM SIGCOMM},
  pages     = {337--348},
  year      = {2007},
  publisher = {ACM}
}

@inproceedings{waldspurger-vmware,
  author    = {Waldspurger, Carl A.},
  title     = {Memory Resource Management in {VMware ESX Server}},
  booktitle = {Proceedings of the 5th USENIX Symposium on Operating Systems Design and Implementation (OSDI)},
  year      = {2002},
  pages     = {181--194},
  address   = {Boston, MA},
  publisher = {USENIX Association}
}

@inproceedings{semantically-smart,
  author    = {Sivathanu, Muthian and Prabhakaran, Vijayan and Popovici, Florentina I. and Denehy, Timothy E. and Arpaci-Dusseau, Andrea C. and Arpaci-Dusseau, Remzi H.},
  title     = {Semantically-Smart Disk Systems},
  booktitle = {2nd USENIX Conference on File and Storage Technologies (FAST)},
  year      = {2003},
  pages     = {73--88},
  address   = {San Francisco, CA},
  publisher = {USENIX Association}
}

@inproceedings{lottery-scheduling,
  author    = {Waldspurger, Carl A. and Weihl, William E.},
  title     = {Lottery Scheduling: Flexible Proportional-Share Resource Management},
  booktitle = {Proceedings of the 1st USENIX Symposium on Operating Systems Design and Implementation (OSDI)},
  year      = {1994},
  pages     = {1--11},
  publisher = {USENIX Association}
}

@inproceedings{facade,
  author    = {Lumb, Christopher R. and Merchant, Arif and Alvarez, Guillermo A.},
  title     = {Fa\c{c}ade: Virtual Storage Devices with Performance Guarantees},
  booktitle = {2nd USENIX Conference on File and Storage Technologies (FAST)},
  year      = {2003},
  pages     = {131--144},
  publisher = {USENIX Association}
}

@inproceedings{triage,
  author    = {Karlsson, Magnus and Karamanolis, Christos and Zhu, Xiaoyun},
  title     = {Triage: Performance Differentiation for Storage Systems Using Adaptive Control},
  booktitle = {Proceedings of the 3rd USENIX Conference on File and Storage Technologies (FAST)},
  year      = {2004},
  pages     = {175--188},
  publisher = {USENIX Association}
}

@inproceedings{pisces-storage,
  author    = {Shue, David and Freedman, Michael J. and Shaikh, Anees},
  title     = {Performance Isolation and Fairness for Multi-Tenant Cloud Storage},
  booktitle = {Proceedings of the 10th USENIX Symposium on Operating Systems Design and Implementation (OSDI)},
  year      = {2012},
  pages     = {349--362},
  publisher = {USENIX Association}
}

@inproceedings{reflex,
  author    = {Klimovic, Ana and Litz, Heiner and Kozyrakis, Christos},
  title     = {{ReFlex}: Remote Flash $\approx$ Local Flash},
  booktitle = {Proceedings of the 22nd International Conference on Architectural Support for Programming Languages and Operating Systems (ASPLOS)},
  year      = {2017},
  pages     = {345--359},
  publisher = {ACM}
}

@inproceedings{borg,
  author    = {Verma, Abhishek and Pedrosa, Luis and Korupolu, Madhukar and Oppenheimer, David and Tune, Eric and Wilkes, John},
  title     = {Large-scale Cluster Management at {Google} with {Borg}},
  booktitle = {Proceedings of the 10th European Conference on Computer Systems (EuroSys)},
  year      = {2015},
  pages     = {1--17},
  publisher = {ACM}
}

@inproceedings{heracles,
  author    = {Lo, David and Cheng, Liqun and Govindaraju, Rama and Ranganathan, Parthasarathy and Kozyrakis, Christos},
  title     = {Heracles: Improving Resource Efficiency at Scale},
  booktitle = {Proceedings of the 42nd Annual International Symposium on Computer Architecture (ISCA)},
  year      = {2015},
  pages     = {450--462},
  publisher = {ACM}
}

@misc{exascale,
  author    = {{Oracle Corporation}},
  title     = {Oracle Exascale: Intelligent Storage for the Cloud},
  year      = {2025},
  howpublished = {Oracle Technical White Paper},
  note      = {\url{https://www.oracle.com/cloud/exascale/}}
}

@misc{morgan-x11m,
  author       = {Morgan, Timothy Prickett},
  title        = {Oracle Revs Up {Exadata} Database Machines To {X11M}},
  year         = {2025},
  month        = jan,
  howpublished = {The Next Platform},
  note         = {\url{https://www.nextplatform.com/2025/01/10/oracle-revs-up-exadata-database-machines-to-x11m/}}
}

@misc{cube-exadata,
  author       = {{theCUBE Research}},
  title        = {The Compelling Economic Value of Incomparable Database Performance},
  year         = {2023},
  month        = jun,
  howpublished = {Industry Analyst Report},
  note         = {\url{https://thecuberesearch.com/compelling-economic-value-incomparable-database-performance/}}
}

@misc{cw-exascale,
  author       = {{Computer Weekly}},
  title        = {Oracle {Exadata Exascale}: Big Architecture Shift for Small Workloads},
  year         = {2024},
  howpublished = {Computer Weekly},
  note         = {\url{https://www.computerweekly.com/news/366602449/Oracle-Exadata-Exascale-Big-architecture-shift-for-small-workloads}}
}

@misc{exadata-sod,
  author       = {{Oracle Corporation}},
  title        = {Oracle {Exadata} Statement of Direction},
  year         = {2025},
  howpublished = {Oracle Technical White Paper},
  note         = {\url{https://www.oracle.com/a/tech/docs/database/exadata-statementofdirection.pdf}}
}

@misc{sunrise-migration,
  author       = {{DatacenterDynamics}},
  title        = {Swiss Telco Sunrise to Migrate 1{,}000 Databases to {Oracle Exadata Cloud@Customer}},
  year         = {2023},
  howpublished = {DatacenterDynamics News},
  note         = {\url{https://www.datacenterdynamics.com/en/news/swiss-telco-sunrise-to-migrate-1000-databases-to-oracle-exadata-cloudcustomer/}}
}

@misc{appsruntheworld-exacc,
  author       = {{Apps Run The World}},
  title        = {Customers of {Oracle Database Exadata Cloud@Customer}},
  year         = {2025},
  howpublished = {Apps Run The World Customer Database},
  note         = {\url{https://www.appsruntheworld.com/customers-database/products/view/oracle-database-exadata-cloudcustomer}}
}

@misc{appsruntheworld-exadata,
  author       = {{Apps Run The World}},
  title        = {Customers of {Oracle Exadata}},
  year         = {2025},
  howpublished = {Apps Run The World Customer Database},
  note         = {\url{https://www.appsruntheworld.com/customers-database/products/view/oracle-exadata}}
}

@misc{trendforce-nand-2026,
  author       = {{TrendForce}},
  title        = {Memory Makers Prioritize Server Applications, Driving Across-the-Board Price Increases in {1Q26}},
  year         = {2026},
  month        = jan,
  howpublished = {TrendForce Press Center},
  note         = {\url{https://www.trendforce.com/presscenter/news/20260105-12860.html}}
}

\end{document}